# Determination of critical success factors affecting mobile learning: a meta-analysis approach

Muasaad Alrasheedi and Luiz Fernando Capretz
Department of Electrical and Computer Engineering
Western University
London, Ontario, Canada
{malrash, lcapretz}@uwo.ca

## ABSTRACT

With rapid technological advancements, mobile learning (m-Learning) offers incredible opportunities, especially in the area of higher education. However, while interest in this area has been significant and several pilot studies have been conducted within universities, relatively less is known about how higher educational institutions can make efficient use of the m-Learning platform to support teaching and learning. Although there are numerous studies in the area, the lack of this insight is mostly due to the fact that very little effort has been made to collate these studies and determine a common set of key success factors that affect the acceptance of m-Learning within universities. This study conducts a systematic analysis of several studies conducted in the area of m-Learning to assess the critical success factors, by making use of the meta-analysis technique. Our investigation has shown that the most important perceived advantages of m-Learning, from learner perspectives, are collaboration during studies, the prospect of ubiquitous learning in space and time, and user friendly application design.

## INTRODUCTION

The work described in this paper builds on previous work (Alrasheedi & Capretz, 2013a) carried out by the authors, which investigates the critical success factors (CSFs) affecting m-Learning platforms. The mobile phone industry has experienced the fastest rate of growth universally throughout the world. While the technology itself came into prominence in the 1980s, the use of the mobile phone was limited to only about 30% of the global population in 2004 (Paul & Seth, 2012). This figure has increased drastically, and, according to a World Bank estimative, more than 90% of the global population is within the range of a cell phone tower. The number of global subscribers have increased from fewer than 700 million in 2000 to more than five billion in 2010, which was about 70% of the population in this year (The World Bank Institute, 2012). The statistics not only point towards the immense success of the technology, but also highlight the versatility of the mobile phone. The rapid acceptance of the technology only serves to underscore the fact that people are aware of the multitude of benefits of the technology and are interested in using it in their daily lives. The continual addition of sophisticated features has only enhanced the usability of mobile phones in several different application areas.

With the rapid rate of advances in mobile phone technology, hi-tech capabilities are now on hand as educational aids and services for both learners (students) and educators. This has led to the growing prominence of m-Learning, with several pilot projects being set up in universities to demonstrate the technical feasibility and pedagogic possibility in the tertiary education section (Zeng & Luyegu, 2011). The reason for the specific interest in the use of m-Learning in higher education, specifically in the engineering and technology field, is because learners are considered to be sufficiently old and technically competent to understand and exploit the mobile phone interface for educational purposes. Further, most of the technical students in this age group already own mobile phones and thoroughly understand their use (Tsai et al., 2005). Studies have shown learners to be completely in favor of using m-Learning as a learning platform as they believe that this will enhance their educational experience. While there are certain concerns regarding the price of inclusion of the technology, most learners consider it to be a good idea. According to learners, the most attractive feature of m-Learning is the possibility of self-learning at their own pace, place, and time (Vate-U-Lan, 2008).

All these factors should mean that the rate of adoption of m-Learning platforms in universities should be at least on a scale similar to its overall growth. Statistics show that this is not the case. Campuses have been relatively slow to adopt m-Learning as a mainstream platform. For instance, the 2010 Campus Computing survey showed that only 13.1% of higher educational institutions have already developed or enabled m-Learning (Quinn, 2011).

The discrepancy between distribution of mobile devices and their use in higher educational institutions is a very interesting and relevant one. The growing interest in the field has compelled many researchers to scientifically study the m-Learning phenomenon. Despite this, relatively little is known about the big picture of how universities can use mobile devices to support teaching and learning (Zeng & Luyegu, 2011). This is because most of the research in the area of m-Learning is highly subjective and contextual, tailored to the requirements of a specific educational institution. Additionally, the definitions used by researchers to describe what they mean by m-Learning are also different. This makes it difficult to collate various studies in the area of m-learning (Väätäjä et al., 2009).

However, we contend that, despite the highly contextual nature of m-Learning studies, several characteristics are similar and the results could be developed into a framework for assessment of the success of m-Learning. One such framework was presented by Ali et al. (2012), where learning contexts, learning experiences, and design aspects were used to assess the success of m-Learning (Ali et al., 2012). Our research uses the meta-analysis approach to conduct a systematic literature review to determine the CSFs for the success of m-Learning in higher educational institutions.

The paper starts with an insight into the key benefits of m-Learning. This is followed by a brief discussion of the CSFs determined by an historical study. The next section discusses our methodology of meta-analysis to determine the CSFs based on recent studies. This is followed by a discussion of the results of the meta-analysis. The paper concludes with a summary of the conclusions and future implications of this study.

## MOBILE LEARNING

The unique feature of the m-Learning platform that makes it a new educational platform is mobility. The concept of mobility refers to the prospect of having flexibility in terms of time, place, pace, and space that cannot be achieved when using non-mobile versions of devices (Andrews et al., 2010). In theory, m-Learning offers learners the opportunity of learning anytime and everywhere. However, it must be understood that the terms 'anytime and everywhere' are limited from being universally true due to connectivity as well as safety restrictions (Saccol et al., 2010).

Mobility is, however, not the only advantage users receive. A key benefit offered by the m-Learning concept is the feature of collaborative learning. While collaboration is also a part of education in traditional learning scenarios, the use of mobile devices means that learners can now interact with fellow students and educators from different locations even when they are not in a formal classroom. Mobility combined with collaborative learning makes the m-Learning platform different from any other existing learning platform, whether it is traditional face-to-face learning or other technology-based platforms like e-Learning (Kukulska-Hulme & Taxler, 2007).

The mobility feature of the m-Learning platform has several implications and applications. It allows learners to manage the content, scope, and space of their learning. Learners also have control over the time and place where they access learning materials. Professionals use this feature of m-Learning for just-in-time learning. This means that employees learn a particular concept as and when they require it and apply it immediately after learning, instead of following the traditional learning process where they learn at a workshop, store this knowledge in their minds, and then use this information practically at a later date. Last, but certainly not least, the mobility feature allows learners from geographically remote locations to be included as a part of the mainstream educational process, without having to shift their location (Saccol et al., 2010). Thus, the concept of mobility is not limited to students being mobile, but the instructors and learning content are also not tied to a particular location. The mobility of learning content can also be translated into a reduction in processing time and a lack of boundaries to physical access. As can be seen, the mobility of learning content is truly revolutionary and pushes the envelope in the context of learning mechanisms as well as information access (Moura & Carvalho, 2010).

However, despite the increase in mobile usage, especially among college-going students, and the multitude of benefits the platform offers, its adoption into mainstream education has been slow. Many analysts attribute this anomaly to a lack of understanding by educators of how to use the technology to enhance the learning process. University management is also said to be extremely conservative and is reluctant to make large investments and revolutionize their tried and tested mode of imparting education. Not many are impressed even by the documented proof from various research studies showing positive inputs from students and other educators regarding the use of m-Learning. University management is also apprehensive of the impact of rapidly changing technology as well as the issues of security and privacy (Wilen-Daungenti, 2008).

It can be seen that instead of presenting multiple research reports from different universities, a common framework of assessment would be of more interest to university management. This framework can be used to assess the barriers to m-Learning within their own educational institution and the progress can be reviewed periodically to assess its success. Needless to say, the development of such a framework requires a comprehensive knowledge of the critical parameters that affect m-Learning. The present paper is an attempt to collate multiple research studies to arrive at these CSFs.

## CRITICAL SUCCESS FACTORS OF M-LEARNING (REVIEW OF PREVIOUS STUDIES)

The m-Learning platform has changed the learning paradigm, and it has the potential to alter the way education is imparted. Most of the pilot studies reviewing the adoption and success of m-Learning within universities tend to focus only on the technical capabilities. As m-Learning technology is entirely dependent on the interaction between humans and machines, focusing solely on the capability of mobile devices and applications only limits the picture. The perspective of success factors must also extend to the usage of m-Learning in different contexts in addition to user experiences from the points of view of learners, educators, and university management (Andrews et al., 2010).

While several researchers have conducted a study of m-Learning projects for determining CSFs, very little effort has actually been put into collating these studies and coming up with a common set of success factors. Cochrane and Bateman (2010a) are responsible for a handful of recent studies involving a cumulative assessment of CSFs from multiple m-Learning studies. They examined 12 m-Learning intervention studies conducted between 2006 and 2009 and pointed towards a single CSF – pedagogical integration of technology into course criteria and assessment. The researchers agreed that there were several other success factors, though they do not measure the extent to which each factor influences the success of m-Learning in the tertiary education sector (Cochrane Bateman, 2010b). This 2010 study was further limited because, while the individual studies considered m-Learning in a different context, they were limited to the use of mobile web 2.0 in tertiary education (Cochrane, 2010). Cochrane's study also addressed the CSFs but this time the study was limited to the analysis of the application of mobile web 2.0 in tertiary education (Cochrane, 2014).

Another study that evaluated the CSFs for m-Learning was published in 2006. This study was conducted by Naismith and Corlett (2006) and involved an exhaustive study of the literature pertaining to m-Learning, published at various m-Learning conferences between 2002 and 2005 (Naismith & Corlett, 2006). The researchers found that while other studies have found a wide array of factors responsible for the success or failure of m-Learning projects, five of the factors were a part of every m-Learning literature – technology availability, support of the concerned institution, network connectivity, assimilation with study curriculum, student experience, or real life, and technology ownership by learners (Adeyeye et al., 2013).

While the study is detailed in its analysis it has the following two drawbacks that prevent it from being relevant in the present-day context. First, the study itself is more than six years old. As previously discussed, during this period the penetration rates of mobile phones have exploded. People all over the globe, including from remote areas and communities, now have access to mobile phones. Mobile phone features have also become extremely sophisticated during this period, especially with the introduction of the Apple iPhone series and all different brands that offer touch screen smartphones. People are also getting used to the rapid technology advancement in mobile phone technology and a large proportion have already jumped onto the smartphone bandwagon. This means that there is a critical need to re-evaluate the CSFs in light of the present state of adoption of mobile technology among the general population. Furthermore, with each new development in technology there are several other new factors which can influence a person in making personal choices, especially when doing so affects other people (Capuruço Capretz, 2010). The previous study also has another drawback. While the study itself uses information from multiple research studies conducted during a three-year period, it does not make use of any systematic method for analysis nor does it measure the extent to which each factor influences the success of m-Learning in tertiary education. The present study attempts to overcome these drawbacks.

Another recent effort made to determine the CSFs of m-Learning was made by UNESCO. A recent report by UNESCO on m-Learning considers the following factors as essential conditions for successful adoption of m-Learning: affordability, leadership, content, support from educators and parents, well-defined m-learning goals, recognition of informal learning, and defined target learner groups for m-Learning (UNESCO, 2011). Interestingly, the UNESCO study also argues that the specific set of CSFs changes as does the learning environment. The study, while giving valuable insights, is not highly relevant to the present context. As UNESCO says, the success factors change as per context, and their report has considered m-Learning in general – including within schools and universities (UNESCO, 2011). The present study addresses the issue of factors of m-Learning in higher education. At this level, the issues of privacy, responsible use of mobile devices, and technical competence and maturity of students are higher than lower-level students. Hence, the m-Learning context and application design is entirely different. Further, the study is limited to different countries in Asia and Africa (UNESCO, 2011), whereas the present study considers the issue of m-Learning in higher education globally.

**META-ANALYSIS OF CRITICAL SUCCESS FACTORS**

Meta-analysis is basically a systematic literature review using quantitative means. This is different from the traditional literature review, where the analysis is arbitrary, theoretical, and, hence, highly subjective. Several different statistical procedures can be used during meta-analysis of existing studies. The only requirement is that these studies also have similar statistical findings as a result of investigation into the same or similar research questions (Booth et al., 2012).

Preliminary and partial results of this study were published and presented at the IEEE International Conference on Teaching, Assessment and Learning for Engineering (TALE2013) (Alrasheedi & Capretz, 2013a). The results of a meta-analysis conducted to evaluate the CSFs affecting m-Learning. One of the main barriers to conducting a systematic statistical analysis of literature into existing m-Learning studies is that the research questions addressed are highly subjective and examine a variety of the implications of m-Learning. Further, as none of the studies have actually prioritized the factors based on the extent of their influence to the success of m-Learning in higher education, there is no master list that can be used for identifying the presence of CSFs. To overcome this limitation, the present study makes use of all CSFs that came out of existing studies in the area of m-Learning, as discussed in the previous section. These form the basic variables for the meta-analysis. The first step of the meta-analysis study was to detect the presence of these variables in all the studies conducted by researchers on the existing m-Learning projects across the world.

The conditions for the present meta-analysis were:

- The studies must pertain to m-Learning in higher education.
- The studies must have been published in the last 6 years; the cut-off year is 2007. Studies published prior to this year are not included in the analysis.
- The studies must be quantitative, i.e., CSFs have been determined by making use of quantitative analysis methods.
- The description of quantitative analysis used in the study is clear (a few studies were discarded because the quantitative analysis used was arbitrary).

We found a total of 19 studies that satisfied the conditions for meta-analysis. In order to make the references to these studies easier, each study has been assigned a unique Roman numeral, alphabetically arranged in increasing order by date as in the most recent. The numerical list is shown in Table 1.

**Table 1**: Reference list of studies used during meta-analysis.

| *Author References* | *Reference Number* | *Author References* | *Reference Number* |
|---|---|---|---|
| (Liaw & Huang, 2011) | **I** | (Seliaman & Al-Turki, 2012) | **XI** |
| (Cochrane, 2010) | **II** | (Bruck et al., 2012) | **XII** |
| (Hamdeh & Hamdan, 2010) | **III** | (Motiwalla & Bruck, 2013) | **XIII** |
| (Huang & Lin, 2007) | **IV** | (Wand et al., 2009) | **XIV** |

| Author References | Reference Number | Author References | Reference Number |
|---|---|---|---|
| (Özdoğan et al., 2012) | **V** | (Cochrane & Bateman, 2010) | **XV** |
| (Valk et al., 2010) | **VI** | (Cheon et al., 2012) | **XVI** |
| (Scornavacca et al., 2009) | **VII** | (Liu et al., 2010) | **XVII** |
| (Wu et al., 2012) | **VIII** | (Ju et al., 2007) | **XVIII** |
| (Alzaza & Yaakub, 2011) | **IX** | (Chanchary & Islam, 2011) | **XIX** |
| (Phuangthong & Malisawan, 2005) | **X** | | |

As mentioned, the source of the CSFs was a combination of previous studies as well as each of the 19 studies included in the analysis. The next step in the meta-analysis involved detecting the presence of various CSFs in each of the 19 studies and noting their presence and information. A total of 21 factors were discovered in the studies that researchers considered to be important for the success of m-Learning in higher educational institutions. Not all factors were, however, present in all the studies. Table 2 below shows the CSFs and the particular studies where these factors were considered to be important by the researchers (using the references from Table 1 above), and the total number of studies in which the CSFs were among the ones analyzed. This exercise was performed to understand the relative important of each factor.

Further, it is important to understand that the absence of a factor does not mean that it is less important; it means that the researchers (based on their observations and taking into consideration their specific context) have not considered the factor to be either applicable or important. Table 2 shows that factors such as institutional support, technical competence of instructors, and developed assessment techniques were considered only in a single study each. In contrast, some factors, such as ownership, have been cited in more than 10 studies. This disparity has crucial implications when the analysis is conducted using statistical means, as it will skew the comparative analysis. Hence, it is important to make the studies more balanced, which is why it was necessary to remove some of the studies from the eventual meta-analysis. This is the next step of the meta-analysis.

**Table 2**: Presence of CSFs in various studies.

| CSFs | Appearance in Various Studies | |
|---|---|---|
| | *Author References* | *No. of Citations* |
| Availability | I, II, III | 3 |
| *Accesibility* | *II, III, IV, V, VI, VII, VIII, IX* | *8* |
| Affordability | II, VII | 2 |
| Internet Access | VI, IX, X, XI | 4 |
| Connectivity | III, V, IX | 3 |
| *Choice of Mobile Devices* | *II, VIII, IX, XII, XIII, XIV, XV* | *7* |
| Web 2.0 Software | II, XII, XV | 3 |
| *Cross Platform Capability* | *I, III, IX, XII, XIII, XIV, XV* | *7* |
| *Ownership* | *I, III, IV, VII, IX, X, XII, XIII, XIV, XV, XVI, XVII, XVIII, XIX* | *14* |
| Institutional Support | XV | 1 |

| CSFs | Appearance in Various Studies | |
|------|------|------|
| | *Author References* | *No. of Citations* |
| *Content* | *I, III, V, VI, XIII, XVII* | *6* |
| Assimilation with Curriculum | III, VI, VII | 3 |
| Educator Perspectives | II, VII, XV | 3 |
| *Learner Perceptions* | *I, II, III, IV, V, VII, IX, XII, XIII, XIV, XV, XVI, XVII, XVIII* | *15* |
| *Learning Community Development* | *I, II, III, V, VI, IX, XIV, XV* | *8* |
| Develop Assessment techniques | II | 1 |
| Faculty Commitment | II, VI, VII, XV, XVI | 5 |
| User Feedback | II, IV, VI, VII, XV | 5 |
| Technical Competence of Instructors | VI | 1 |
| *Technical Competence of Students* | *I, II, IV, VI, IX, XI, XII, XIII, XIV, XV, XVII* | *11* |
| *User Friendly Design of Content* | *I, II, III, IV, X, XIII, XIV, XVI, XVII, XVIII* | *11* |

Based on the paper by Teoh (2011), the variables, i.e., the CSFs, can be divided into four categories – technology, management support, teaching pedagogy, and learning approach. As seen in Table 3 below, some factors, such as assimilation with curriculum, fall into multiple categories meaning that there is an overlap based on the categories of people influencing the particular variable. The CSFs can be further divided into three main categories – from student perspectives, from instructor perspectives, and from management perspectives.

**Table 3**: Classification of CSFs.

| Variables | CSF Categories |
|-----------|----------------|
| Availability | |
| Accessibility | |
| Affordability | |
| Internet access | Technology |
| Connectivity | |
| Choice of Mobile Devices | |
| Web 2.0 software | |
| Cross-platform capability | |
| Ownership | |
| Institutional Support | |
| Administrative support | Management Support |
| Assimilation with Curriculum | |
| User feedback | |
| Educator perceptions | |
| Technical competence of instructors | |
| Faculty commitment | Teaching Pedagogy |
| Develop assessment techniques | |
| User feedback | |
| Assimilation with Curriculum | |
| Learning community development | |
| User feedback | |
| Learner perceptions | |
| Technical competence of students | Learning Approach |
| User friendly design of content | |
| Assimilation with Curriculum | |

An interesting part of the analysis was that learner perceptions was included in every single study and all of the studies had evaluated the CSFs from this perspective. Any m-Learning platform had two other sets of users – the instructors, and university management and administration. These have not been the focus of evaluative studies, which is a major drawback of the present study. Coming back to the present evaluative study, it can be seen that most of the pilot studies conducted in the area of m-Learning tend to evaluate the success from the perspective of learners, in other words, the factors that are important for learners to ensure that m-Learning is successfully adopted within a tertiary education institution. This is the tentative choice for the independent variable (highlighted in bold in Table1).

As discussed before, the next step involves removing the CSFs that have appeared in very few studies so that only significant CSFs are considered. The removal of the studies has been conducted by using a threshold, i.e., the minimum number of studies in which a factor has to appear before it is included in the meta-analysis. This threshold was chosen to be six, which is a little less than half of the maximum number of appearances (learner perceptions – 15). Applying the threshold results in a total of nine CSFs that have been plotted against the number of citations in Figure 1.

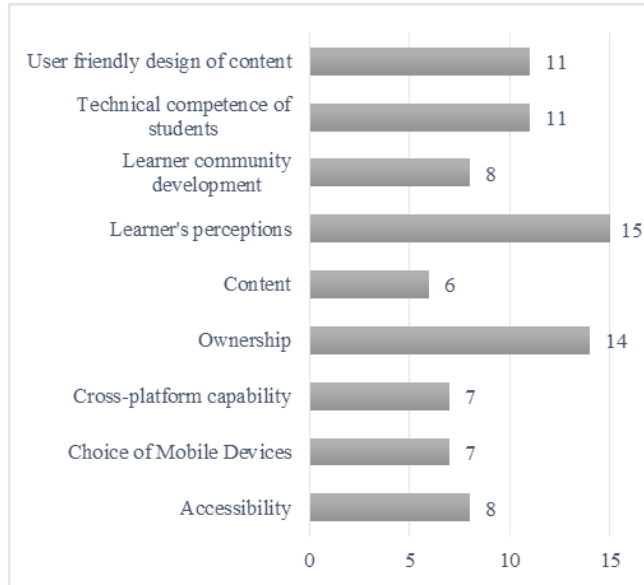

Figure 1. Plot of Shortlisted CSFs against the number of citations.

From Figure 1 above, the nine shortlisted critical success factors are – user friendly design, technical competence, learner community development, learner perceptions, content, ownership, accessibility, choice of mobile device, and cross platform capability.

The next step of the meta-analysis involved conducting in-depth studies of the statistical information available on these nine shortlisted CSFs. The purpose was to understand whether there was sufficient information for each of the factors to conduct a meta-analysis using statistical techniques. The study showed that not enough statistical data was available for three of the CSFs – accessibility, choice of mobile device, and cross platform capability. Hence, these three factors had to be excluded from the final list. This resulted in a total of six CSFs being used for the purpose of meta-analysis (shown in italics, both regular and bold fonts, in Table 4 below).

At this point, it is important to point out that the fact that only six of the 21 CSFs were shortlisted means that these six factors are considered by all researchers studying m-Learning to be important. The remaining factors may be important too, but corroborating their importance would require researchers to include these factors as a part of the study. These six factors – user friendly design, technical competence, learner community development, learner perceptions, content, and ownership – would not be used for a cross-sectional analysis across multiple studies.

The next step of the meta-analysis required shifting focus back to the papers to find out if sufficient statistical information was available for analysis. In other words, the six critical factors must be presented in a similar or at least inter-convertible statistical form so that they can be assessed on a comparative scale. In addition to determining the CSF, one of the objectives of the present study was also to determine the influence each factor has on the success of m-Learning in higher educational institutions. This requires comparison among the CSFs and, hence, this exercise. Our analysis showed that only nine of the 18 studies had similar statistical information that could be used for conducting meta-analysis – I, V, VII, IX, XI, XIII, XVI, and XVIII. This also means that the remaining 10 studies – II, III, IV, VI, VIII, X, XII, XIV, XV, and XIX – were discarded as they either used very different statistical measurements or did not have sufficient raw data required for analysis. The six CSFs and their corresponding statistics in the nine studies are enumerated in Table 4 below.

**Table 4:** CSF statistics in the shortlisted studies.

| Statistics | Critical Success Factors | | | | | |
| --- | --- | --- | --- | --- | --- | --- |
| | *User Friendly Design* | *Technical Competence* | *Learner Community Development* | *Learner Perceptions* | *Content* | *Ownership* |
| Based on (Liaw & Huang, 2011) No of Participants 168 | | | | | | |
| SD | 3.63 | 2.74 | 4.03 | 3.43 | 4.09 | 3.14 |
| Mean | 1.39 | 1.65 | 1.39 | 1.53 | 1.24 | 1.57 |
| Based on (Özdoğan, Başoğlu, & Erçetin, 2012) No of Participants 81 | | | | | | |
| SD | 4.27 | 4.05 | 3.4 | 3.85 | 3.63 | 3.95 |
| Mean | 0.97 | 1.18 | 1.37 | 1.01 | 1.32 | 1.04 |
| Based on (Scornavacca, Huff, & Marshall, 2009) No of Participants 569 | | | | | | |
| SD | 4.04 | Na | 4.05 | 3.71 | 2.95 | 3.67 |
| Mean | 1 | Na | 1.02 | 1.06 | 1.01 | 1.1 |
| Based on (Alzaza & Yaakub, 2011) No of Participants 261 | | | | | | |
| SD | 4.05 | Na | 3.91 | 3.87 | 4.5 | Na |
| Mean | 0.63 | Na | 0.66 | 0.76 | 0.81 | Na |
| Based on (Seliaman & Al-Turki, 2012) No of Participants 55 | | | | | | |
| SD | Na | 4.02 | Na | 4.12 | Na | 4.01 |
| Mean | Na | 1.06 | Na | 0.66 | Na | 0.71 |
| Based on (Motiwalla & Bruck, 2013) No of Participants 33 | | | | | | |
| SD | 3.3 | 3.11 | Na | 2.43 | 2.49 | 3.97 |
| Mean | 0.79 | 1.02 | Na | 1.15 | 1.16 | 1.07 |
| Based on (Cheon, Lee, Crooks, & Song, 2012) No of Participants 177 | | | | | | |
| SD | 3.72 | 3.9 | Na | 3.48 | 3.71 | 3.71 |
| Mean | 1.44 | 1.14 | Na | 1.44 | 1.29 | 1.35 |
| Based on (Liu, Liu, & Carlsson, 2010) No of Participants 219 | | | | | | |
| SD | 3.8 | Na | Na | 3.33 | Na | 3.31 |
| Mean | 1.24 | Na | Na | 1.3 | Na | 1.31 |
| Based on (Ju, Sriprapaipong, & Minh, 2007) No of Participants 245 | | | | | | |
| SD | 2.36 | Na | Na | 2.19 | 2.38 | 2.23 |
| Mean | 1.06 | Na | Na | 1.16 | 1.12 | 1.16 |

From Table 4 above it is clear that the CSF learner perceptions is present in all the nine studies shortlisted above. As discussed earlier, learner perceptions were present in the maximum number of studies. The presence of learner perceptions in all nine studies indicates that learner perceptions actually refers to whether the learners would consider opting for m-Learning in the future based on their current experiences. As the success of m-Learning directly refers to the continued usage of an m-Learning platform, this CSF becomes even more important. In fact, in several studies, learner perceptions were actually correlated with other CSFs as a means for judging the success of mobile l-Learning in a particular institution.

The present meta-analysis also uses learner perceptions as a means of assessing the success of m-Learning in various institutions, i.e., as a dependent variable. The individual correlations are not available for some of the nine studies, hence the meta-analysis consisted of aggregating the mean values of the remaining five CSFs for these studies and then correlating them with learner perceptions. Microsoft Excel was used as a means of performing this operation. The meta-analysis conducted by Ravesteyn and Batenburg is the basis for the present study (Ravesteyn & Batenburg, 2010). The meta-analysis results are shown in Table 5 below.

**Table 5**: Meta-analysis of CSF Statistics.

| CSFs | Meta-Analysis Statistics | | | | | |
|---|---|---|---|---|---|---|
| | *No. of Studies* | *No. of Participants* | *Net Mean* | *Net SD* | *CSF Rank* | *Pearson Corr.* |
| Learner Perceptions | 9 | 1808 | 3.379 | 1.119 | NA | 1 |
| User Friendly Design | 8 | 1753 | 3.646 | 1.065 | 3 | 0.92961 |
| Learner Community Development | 5 | 514 | 3.564 | 1.21 | 4 | 0.64153 |
| Technical Competence | 4 | 1079 | 3.848 | 1.11 | 2 | -0.5595 |
| Content | 7 | 1289 | 3.958 | 1.136 | 1 | 0.80454 |
| Ownership | 8 | 1547 | 3.499 | 1.164 | 5 | 0.6064 |

**RESULTS AND DISCUSSION**

The success of m-Learning is dependent on the views of the users of the m-Learning platform. The popularity of mobile phones in the present day world cannot be denied; neither can their increased invasion into all aspects of people's lives. Despite this, the use of mobile technology in the educational sector has been limited. Consequently, higher educational institutions have recently been looking at several methods of implementing m-Learning strategies (Alrasheedi & Capretz, 2013b). As the popularity and the all-encompassing nature can only come through a favorable user perception, it can be concluded that the users have certain reservations when it comes to the use of mobile technology in the educational sector. The objective of this paper is to assess user perceptions of what users consider to be the key factor necessary for the successful adoption of m-Learning in educational institutions. We conducted a meta-analysis of existing studies that evaluated the CSFs of the m-Learning platform. The results of the analysis are given in Table 5 above.

Table 4 shows the aggregated results from the meta-analysis of nine similar studies conducted measuring the CSFs of m-Learning. The independent observation of the means shows the response tendency on a 5-point Likert scale (1 – strongly disagree to 5 – strongly agree). This shows that a mean over 2.5 indicates that learners agree that the factor has an appreciable influence on their current experience with the m-Learning platform. From Table 5, it can be seen that all six CSFs have an aggregate response higher than 2.5, indicating that each of these factors has an appreciable influence on their current experience with the m-Learning platform.

The rankings of the means of the responses, given in Table 5, show how much influence the factor has on the potential success of m-Learning, according to the learners. It is seen that content is considered to have the most influence, followed by technical competence of learners, user friendly design, learner community development, and ownership.

The next step is to find if the factors are seen to positively affect learner perceptions of m-Learning. As discussed earlier, correlation of the CSFs with learner perceptions is a means of judging the success factors of m-Learning from learner perspectives. Both content and user friendly design have highly positive correlations with learner perceptions. This means that both good content and user friendly design of the content are essential to learners if they are to choose an m-Learning platform in the future. Ownership, i.e., flexibility to use m-Learning anytime and anyplace, and learner community development, i.e., using the m-Learning platform to connect with other learners or educators, are also positively related with learner perceptions. This means that learners view both of these factors as important. Interestingly, technical competence is negatively correlated with learner perceptions. Developing the right evaluation framework will enable researchers and m-Learning stakeholders to get a true picture of the current status of m-Learning implementation and adoption within the educational institution; it can then also be used as a roadmap for success that includes adoption of m-Learning at important milestones (Alrasheedi & Capretz, 2013c). This means that learners consider that they already have technical capabilities (since mobile phones are ubiquitous in the present day world), and so the factor is not critically important in their choice for selecting the m-Learning platform in future.

## CONCLUSIONS AND RECOMMENDATIONS

The goal of this study has been to extend our understanding of the CSFs that affect the m-Learning platform in higher educational settings. A meta-analysis of multiple studies was conducted on the lines of the inspiring study by Naismith & Corlett (2006) using updated data, though the version also improves upon the 2006 study by conducting it as a systematic statistical analysis. This would help in assessing the influence of each CSF on the success of m-Learning in higher education.

An important observation made during the study was that usually the success of m-Learning is considered only based on perceived benefits from learner perspectives (students). This neglects other important users such as the instructors and members of the university management and administration. These people are the first in line to accept or reject any new learning paradigm. They are also responsible for motivating university students into trying out the new platform and helping with the glitches. Unless detailed information is available on how users of these categories think, the information available for the success of m-Learning is incomplete. This means that the universities will have to assess the opinions of their students before any m-Learning initiative is begun.

The categorical division of the CSFs gave a general idea about the causes of the success of m-Learning initiatives in different countries. A careful look at Table 3 above shows that the success factors, when examined from the category perspective, are fairly evenly distributed. This means that m-Learning success depends on a set of factors, only one part of which is technical.

The meta-analysis of the literature review also showed that some aspects – such as the technical competence of educators, the development of assessment techniques, and institutional support – have been considered by very few studies as success factors. This does not mean that the factors are not important. It is in fact a possible explanation of the slow adoption of the technology in the educational sector. As a future study, we are working on systematically surveying the CSFs for m-Learning to investigate and classify these CSFs into different groups such as from the perspectives of students, instructors, and university management. This would be promising background for proposing a new conceptual framework to comprehensively study and analyze the relationship among the CSFs from related perspectives.

### ACKNOWLEDGMENT

The first author would like to thank the ministry of higher education (MOHE) in Saudi Arabia for his personal fund and the scholarship.

## REFERENCES

Adeyeye, M. O., Musa, A. G., & Botha, A. (2013). Problem with multi-video format m-learning applications. In J.-E. Pelet, E-Learning 2. 0 Technologies and Web Applications in Higher Education (pp. 294-330). New York: IGI Global.

Ali, A., Ouda, A., & Capretz, L. F. (2012). A conceptual framework for measuring the quality aspects of mobile learning. *Bulletin of the IEEE Technical Committee on Learning technology, 14*(14), 31-34.

Alrasheedi, M. & Capretz, L.F. (2013a). A Meta-Analysis of Critical Success Factors Affecting Mobile Learning, *Proceedings of IEEE International Conference on Teaching, Assessment and Learning for Engineering,* Bali, Indonesia, pp. 262-267.

Alrasheedi, M., & Capretz, L. F. (2013b). Applying CMM towards an m-learning context. In International Conference on Information Society (i-Society), (pp. 146-151). IEEE.

Alrasheedi, M., & Capretz, L. F. (2013c). Developing a Mobile Learning Maturity Model. International Journal for Infonomics (IJI), 6(3/4), 771- 779.

Alzaza, N. S., & Yaakub, A. R. (2011). Students' awareness and requirement of mobile learning in higher education environment. America Journal of Economics an Business Administration, 3(1), 95-100.

Andrews, T., Smyth, R., Tynan, B., Berriman, A., Vale, D., & Cladine, R. (2010). Mobile technologies and rich media: expanding tertiary education opportunities in developing countries. In A. G. Abdel-Wahab, & A. A. El-Masry, Mobile Information Communication Technologies Adoption in Developing Countries: Effects and Implication. New York: Idea Group Inc.

Booth, A., Papaioannou, D., & Sutton, A. (2012). Approaches to successful literature review. California: SAGE.

Bruck, P. A., Motiwalla, L., & Foerster, F. (2012). Mobile elarning with micro-content: a framework and evaluation. 25th Bled e-Conference eDependability: Reliable and Trustworthy eServices for the Future, 527-543.

Capuruço, R. A., & Capretz, L. F. (2010). Integrating recommender information in social ecosystems decisions. In Proceedings of the Fourth European Conference on Software Architecture: Companion Volume (pp. 143-150), ACM.

Chanchary, F. H., & Islam, S. (2011). Mobile learning in Saudi Arabia - prospects and challenges. International Arab Conference on Information Technology (ACIT'2011). Jordan: Zarqa University.

Cheon, J., Lee, S., Crooks, S. M., & Song, J. (2012). An investigation of mobile learning readiness in higher education based on the theory of planned behavior. Computer & Education, 59, 1054-1064.

Cochrane, T. D. (2010). Exploring mobile learning success factors. Research in Learning Technology, 18(2), 133-148.

Cochrane, T. D. (2014). Critical success factors for transforming pedagogy with mobile Web 2.0. British Journal of Educational Technology, 45(1), 65-82.

Cochrane, T. D., & Bateman, R. (2010a). Reflections on 3 years of m-learning implementation (2007-2009). Sheffield: Sheffield Hallam University Research Archive.

Cochrane, T. D., & Bateman, R. (2010b). Transforming pedagogy using mobile web 2.0. In Web-based Education: Concepts, Methodologies, Tools and Applications, Volume 1 (pp. 671-698). New York: Idea Group Inc (IGI).

Hamdeh, M. A., & Hamdan, A. (2010). Using analytical hierarchy process to measure critical success factors of m-learning. European, Mediterranean & Middle Eastern Conference on Information Systems. Abu Dhabi.

Huang, J. H., & Lin, Y. R. (2007). Elucidating user behavior of mobile learning: a perspective of the extended technology acceptance model. The Electronic Library, 25(5), 585-598.

Ju, T. L., Sriprapaipong, W., & Minh, D. N. (2007). On the Success Factors of Mobile Learning. 5th International Conference on ICT and Higher Education. Bangkok.

Kukulska-Hulme, A., & Taxler, J. (2007). Designing for mobile and wireless learning. In H. Beetham, & R. Sharpe, Rethinking Pedagogy for a Digital Age: Designing and Delivering e-Learning (pp. 180-192). London: Roultedge.

Liaw, S. S., & Huang, H. M. (2011). Exploring learners' acceptance toward mobile learning. In T. Teo, Technology Acceptance in Education: Research and Issues (pp. 145-157). Sense Publishers.

Liu, Y., Liu, H., & Carlsson, C. (2010). Factors driving the adoption of m-learning: an empirical study. Computers & Education, 22, 1211-1219.

Motiwalla, L., & Bruck, P. A. (2013). Adoption of a micro-learning system. *Tenth AIMS International Conference on Management,* Austria, 1001-1008.

Moura, A., & Carvalho, A. (2010). Mobile learning: using SMS in educational contexts. In N. Reynolds, & M. Turcsányi-Szabó (Ed.), Key Competencies in the Knowledge Society, World Computer Congress (pp. 281-291). Brisbane: Springer.

Naismith, L., & Corlett, D. (2006). Reflections on success: a retrospective of the mLearn conference series 2002-2005. mLearn 2006 – Across generations and cultures. Banff.

Özdoğan, K. M., Başoğlu, M., & Erçetin, G. (2012). Exploring major determinants of mobile learning adoption. Proceedings of PICMET '12: Technology Management for Emerging Technologies (pp. 1415-1423). IEEE.

Paul, J., & Seth, R. (2012). Japan-India diplomacy and relationship marketing. In Paul, International Marketing – text and Cases (2nd ed., pp. 178-180). New Delhi: Tata McGraw-Hill Education.

Phuangthong, D., & Malisawan, S. (2005). A study of behavioral for 3G mobile internet technology: preliminary researh mobile learning. Proceedings of the Second International Conference on e-Learning for Knowledge-Based Society, (pp. 1-17). Bangkok.

Quinn, C. N. (2011). The mobile academy: mLearning for higher education. San Francisco: John Wiley & Sons.

Ravesteyn, R., & Batenburg, R. (2010). Surveying the critical success factors of BPM-systems implementation. Business Process Management Journal, 3, 492-507.

Saccol, A., Barbosa, J. L., Schlemmer, E., & Rienhard, N. (2010). Corporate m-learning: applications and challenges. In R. Guy, Mobile Learning: Pilot Projects and Initiatives. California: Information Science.

Scornavacca, E., Huff, S., & Marshall, S. (2009). Mobile phones in the class: if you can't beat them, join them. *Communications of the ACM, 52*(4), 142-146.

Seliaman, M. E., & Al-Turki, M. S. (2012). Mobile learning adoption in Saudi Arabia. World Academy of Science, Engineering and Technology, 69, 391-293.

Teoh, K. K. (2011). An examination of critical success factors in the implementation of ePortfolios in universities. *Journal of Academic Language and Learning, 5*(2), A60-A72.

The World Bank Institute. (2012). Behavioral change using technology. Retrieved March 1, 2014, from http://wbi.worldbank.org/wbi/content/behavioral-change-using-technology

Tsai, L. H., Young, S. S., & Liang, C. H. (2005). Exploring the course development model for the mobile learning context: a preliminary study. Fifth IEEE International Conference on Advanced Learning Technologies ICALT05, 5-8.

UNESCO. (2011). UNESCO mobile learning week report. Retrieved March 1, 2014, from http://www.unesco.org/new/fileadmin/MULTIMEDIA/HQ/ED/ICT/pdf/UNESCO%20MLW%20report%20final%2019jan.pdf

Valk, J. H., Rashid, A. T., & Elder, L. (2010). Using mobile phones to improve educational outcomes: an analysis of evidence from Asia. International Review of Research in Open and Distance Learning, 11(1), 117-140.

Vate-U-Lan, P. (2008). Mobile learning: major challenges for engineering education. 38th ASEE/IEEE Frontiers in Education Conference (pp. 11-16).

Väätäjä, H., Männistö, A., Vainio, T., & Jokela, T. (2009). Understanding user experience to support learning for mobile journalist's work. In G. R, The Evolution of Mobile Teaching and Learning (pp. 177-210). California: Informing Science Press.

Wand, Y. S., Wu, M. C., & Wang, H. Y. (2009). Investigating the determinants and age and gender differences in the acceptance of mobile learning. British Journal of Educational Technology, 4(1), 92-118.

Wilen-Daungenti, T. (2008). Edu technology: technology and learning environments in higher education. New York: Peter Lang.

Wu, W. H., Wu, Y. C., Chen, C. Y., Kao, H. Y., Lin, C. H., & Huang, S. H. (2012). Review of trends from mobile learning studies: a meta-analysis. Computers & Education, 59, 817-827.

Zeng, R., & Luyegu, E. (2011). Mobile learning in higher education. In A. D. Olofsson, & J. O. Lindberg, Informed Design of Educational Technologies in Higher Education: Enhanced Learning and Teaching (pp. 292-306). Hershey, Philadelphia: Idea Group Inc (IGI).